\documentstyle[prl,aps,epsf,twocolumn]{revtex}
\newcommand{\eq}{\begin{equation}}
\newcommand{\ee}{\end{equation}}
\newcommand{\s}{{\sigma}}  

\newcommand{\rb}{{\bar{\rho}}}
\newcommand{\phib}{{\bar{\phi}}}

\newcommand{\dva}{{\frac{\vp\times\va}{2\pi}-\rb}}

\newcommand{\w}{{\omega}}
\newcommand{\zh}{{\hat{z}}}

\newcommand{\vA}{{\vec{A}}}
\newcommand{\va}{{\vec{a}}}
\newcommand{\vrr}{{\vec{r}}}
\newcommand{\vj}{{\vec{j}}}

\newcommand{\vx}{{\vec{x}}}
\newcommand{\vq}{{\vec{q}}}      

\newcommand{\vb}{{\vec{b}}}
\newcommand{\vp}{{\vec{\partial}}}
\newcommand{\p}{{\partial}}

\newcommand{\gr}{{\nabla}}
\newcommand{\ra}{{\rightarrow}}
\def\eqa{\begin{eqnarray}}
\def\eea{\end{eqnarray}}
\begin{document}
\draft
\flushbottom
\twocolumn[
\hsize\textwidth\columnwidth\hsize\csname @twocolumnfalse\endcsname
\title{Neutral Fermions at $\nu=1/2$}
\author{Dung-Hai Lee}
\address{Department of Physics, University of California at
Berkeley, Berkeley, CA 94720}

\date{\today}
\maketitle
\tightenlines
\widetext
\advance\leftskip by 57pt
\advance\rightskip by 57pt

\begin{abstract}

We present a particle-hole symmetric theory for a two-dimensional electron gas at filling factor one half. 
In this theory, elementary excitations are 
dipole-like fermions floating on top of the $\nu=1/2$ boson quantum Hall liquid. In the absence of disorder these dipoles can form a Fermi-liquid. Disorder can break them apart and drive the ground state into a quantum critical  state.

\end{abstract}
\vskip 1cm

]

\narrowtext
\tightenlines

The discovery of an acoustic wave anomaly\cite{willet} in quantum Hall regime, marked the begining of the  ``$\nu=1/2$ problem''. To explain the experiment, a novel idea, the composite Fermi liquid theory (CFLT), was proposed by Halperin, Lee and Read\cite{hlr}. These authors represent an electron as a Chern-Simons (CS) fermion carrying two quanta of fictitious magnetic flux.\cite{jain}  At half filling, the average fictitious flux cancels the real one, thus allowing the CS fermions to form a Fermi liquid. CFLT theory asserts that this hidden Fermi liquid is the root of the acoustic wave anomaly observed at $\nu=1/2$. 
This idea obtained wide acceptance when experimental evidence indicating the existence of a Fermi surface was reported.\cite{fsurf} 

Since in CS approach the fictitious flux is dynamic, 
the concept of canceling it against the real, static flux requires further scrutiny. The thesis of CFLT is that dynamic flux fluctuations will just renormalize the parameters of the Fermi liquid. However, when actual calculations are made to corroborate such claim, it is found that the renormalizations diverge.\cite{hlr,stern}

In order to avoid ambiguity, let us define the phrase CFLT. In CS formalism, it is possible to prove that the electron resistivity tensor 
$\rho_{\alpha\beta}$ ($\alpha,\beta = x,y$) can be written as the sum of two terms:
\eq                                 
\rho_{\alpha\beta}=\rho_{\alpha\beta}^{CS}+2\frac{h}
{e^2}\epsilon_{\alpha\beta}.
\label{8989}
\ee
In the above $\rho_{\alpha\beta}^{CS}$ has the property that its inverse $\sigma_{\alpha\beta}^{CS}$ is the sum of all CS fermion particle-hole diagrams that are irreducible with respect to the cutting a fictitious gauge line\cite{hlr,klz}. In the absence of an exact knowledge of $\rho_{\alpha\beta}^{CS}$, CFLT {\it asserts} that it takes a Fermi liquid form.

In a recent paper\cite{lkgk} Lee, Krotov, Gan and Kivelson (LKGK) challenge the above assertion.
In short, they show that when electrons are confined to the lowest Landau level, particle-hole symmetry requires their Hall conductivity to be $e^2/2h$\cite{jiang1}. They further point out that such a value of $\sigma_{xy}$ implies
$\s_{xy}^{CS}=-e^2/2h$ so long as $\rho_{xx}\ne 0$ (or when the degree of disorder is nonzero). Of course a nonzero $\sigma_{xy}^{CS}$ is inconsistent with a Fermi liquid form. Since LKGK's arguments also apply to finite temperature and finite system sizes, one can not view such inconsistency as due to certain infrared instability (such as localization) induced by disorder.

Strong disorder also presents a challenge to CFLT. For example, in the presence of a strong\cite{str} {\it particle-hole symmetric}, disorder potential the ground state will break up into droplets of the $\nu=1$ quantum Hall liquid. Moreover, due to particle-hole symmetry these droplets will critically (quantum) percolate. The challenge for CFLT is to account for such critical state.

The following is a summary of our results.
1) The state at $\nu=1/2$ can be viewed as a liquid of fermion dipoles (neutral fermions) floating on top of the $\nu=1/2$ boson Laughlin liquid. 
2) Each dipole is made up of a pair of charge $\pm1/2$ and statistics $\pi/2$ anyons. 
4) The conductivity tensor ($\s_{\alpha\beta}^N$)
of the dipole liquid is related to that of electrons by
\eq
\s_{\alpha\beta}=\frac{1}{2}\frac{e^2}{h}
\epsilon_{\alpha\beta} +\s_{\alpha\beta}^N.
\label{newcon}
\ee
5) $\s_{\alpha\beta}^N$ is given by
\eqa
&&\s_{xy}^N=0\nonumber \\
&&\s_{ll}^N=\frac{1}{i\w}\frac{\w^2}{(4\pi)^2\Pi_{tt}(\vq,\w)+q^2V(\vq)}\nonumber \\
&&\s_{tt}^N=\frac{1}{i\w}\Pi_{tt}(\vq,\w),
\label{result}
\eea
where $\Pi_{tt}(\vq,\w)$ is the transverse current-current correlation 
function of electrons. 
6) Strong disorder can break the neutral fermion apart, and drive the 
ground state into the critically percolated state discussed above.

To understand the physical implications of $\sigma_{xy}=e^2/2h$, let us imagine inserting a CS fermion at a point in the $\nu=1/2$ liquid. Since a CS fermion is an electron carrying two quanta of fictitious flux, such insertion produces a local surplus of magnetic flux. Through $\s_{xy}$ the EMF of this time-dependent flux induces an radial outward flow of electrons. When $\sigma_{xy}=e^2/2h$, the time-integrated charge that flow out 
is $2\times\frac{1}{2} e=e$. Consequently the charged CS fermion is dressed with a correlation hole and becomes 
neutral. The resulting neutral object is called a ``composite fermion'' in an insightful paper of Read\cite{read}. Aside from the present work, there are currently two other ongoing attempts at describing the composite fermions.\cite{shankar,pasquier} However, the results listed in 1)-6) are all new predictions of the present paper. 

According to our physical picture, an electron is viewed as a CS boson carrying a quantum of fictitious magnetic flux. Were it not for the fictitious flux the CS bosons, 
each experiencing $\nu=1/2$, would condense into the $\nu=1/2$ Laughlin state $\Psi_{1/2}=\prod_{(ij)}(z_i-z_j)^{2} \exp\{-\sum_k|z_k|^2/4l_B^2\}$. Each quantum of fictitious flux induces a charge $1/2$, statistics $\pi/2$, quasihole. In addition, after having been screened by the Laughlin liquid, each fictitious flux quantum behaves as 
a charge $-1/2$, statistics $\pi/2$ quasiparticle itself.
To minimize the electrostatic energy, a quasiparticle binds to a quasihole. The resulting fermionic (since statistics $=\pi/2+\pi/2=\pi$) bound state is what we call a neutral fermion. Despite the lack of obvious relation between our approach and others, we believe that our neutral fermion is the same as the composite fermion in Ref.\cite{read,shankar,pasquier}.

It is also possible to rationalize the above picture from  a  wavefunction consideration. We recall that the wavefunction 
$\Psi=P_{LLL}[det({z_j,\bar{z}_j})\Psi_{1/2}]$ has an almost perfect overlap with the numerical ground state for systems consisting of small numbers of particles.\cite{rezyai}
In the above $P_{LLL}$ is the lowest Landau level projection operator, and $det$ is the Slater determinant describing a Fermi sea with $k_F=\sqrt{4\pi\bar{\rho}}$. The operator $P_{LLL}$ causes $e^{ik_j\bar{z}_j}$ 
in $det$ to become $e^{2ik_j\p/\p z_i}$, i.e., a displacement operator. 
As a result, $P_{LLL}$ shifts the zeroes of $\Psi_{1/2}$.\cite{read}  
Before such shift, all zeroes of $\Psi_{1/2}$ are attached to particles. This will no longer be true after the shift. However, since the final $\Psi$ is a fermion wavefunction in the lowest Landau level, it must be expressible as  
$\Psi=S(\{z_k\})\prod_{(ij)}(z_i-z_j)\exp\{-\sum_k|z_k|^2/4l_B^2\}$, where, for rotational invariant states, $S$ is a symmetric homogeneous 
polynomial of degree $N-1$. (N is the total electron number.) Therefore, the net effect of $P_{LLL}$ is 
to shift away one out of the two zeroes attached to each particle. The final $\Psi$ can be viewed as a wavefunction descrbing $N$ charge $1/2$ statistics $\pi/2$ and 
$N$ charge $-1/2$  statistics $\pi/2$ anyons above the bosonic Laughlin state $\Psi_{1/2}$.
 
At this juncture it is worth noting that a RPA treatment in the CS fermion approach predicts the ground state wavefunction to be $ det \Psi_{1/2}$. Thus the effect of
$P_{LLL}$ (and hence the displacement of wavefunction zeroes) is hidden beyond RPA. It should also be emphasized that the shifting of one of the zeroes of $\Psi_{1/2}$ away from each particle represents a {\it strong, short wavelength} fluctuation of the fictitious magnetic flux.

In the CS formulation, the following action governs the dynamics of the CS boson:
\eqa
&&S=\int d^2xdt\{\phib D_0\phi
+\frac{1}{2m}|(\vec{D}+i\vec{A}_{ext})\phi|^2
+\frac{i}{4\pi}a\cdot\gr\times a\}
\nonumber \\
&&+S_{int}[\delta\rho],\nonumber \\
&&S_{int}=\frac{1}{2}\int dtd^2xd^2x'
V(\vx-\vx')\delta\rho(\vx)\delta\rho(\vx').
\label{1}
\eea
In the above, $\phi$ is the CS boson field; $D_{\mu}\equiv\p_{\mu}-ia_{\mu}+iA_{\mu}$, where $a$ is the statistical gauge field, and $A$ is the probing gauge field; $\delta\rho(\vx)\equiv|\phi(x)|^2-\rb$ where $\rb$ is the average electron density; $V(\vx-\vx')$ is the electron-electron interaction potential;$\vA_{ext}$, satisfying 
$\vp\times\vec{A}_{ext}=B$, is the external vector potential. 
In the above and hereafter symbols without explicit vector sign, such as $a, A, \gr, x$ denote 2+1 dimensional vectors, while those with explicit vector signs, such as $\va,\vA,\vp,\vx$ denote 2-dimensional vectors. In Eq.(\ref{1}) and the rest of the paper we shall
adopt units and rescale gauge fields such that $\hbar=c=e=1$.

In the limit where the bare electron mass vanishes 
(or when the cyclotron energy diverges), the dynamics of  $a$ and the vortices induced by it govern the low energy physics. The effective action\cite{note2} describing them is given by: 
\eqa
S&=&\int d^2xdt
\{-\frac{i}{8\pi}(a-b-A)\cdot\gr\times (a-b-A) \nonumber \\
&+&\frac{i}{4\pi}a\cdot\gr\times a
-ib\cdot j-i\rb(a_0-b_0-A_0)\}\nonumber \\
&+&S_{int}[j_0-\rb].
\label{2}
\eea
In the above $\frac{1}{2\pi}\gr\times\vb$ is the vortex 3-current, and 
\eq
j_0=\sum_j\delta(\vx-\vrr_j),\hspace{0.1in}
\vj=\sum_i\dot{\vrr}_i\delta(\vx-\vrr_i),
\label{dii}
\ee
the 3-current density of point-like unit charges, is introduced to quantize $\vp\times\vb$. (We emphasize that after the projection to the lowest Landau level, even a point-like particle acquires a fuzziness $l_0$ of the order magnetic length.)
>From the first term in Eq.(\ref{2}) we deduce that each quantum of 
$\vp\times\va$ (or $\vp\times\vb$) carries charge $-1/2$ (or $1/2$). 
We note that in Eq.(\ref{2}) the bare electron mass drops out, and the 
parameters in $S_{int}$ set the energy scale.

In the absence of the probing field, we can fix the gauge
($\vp\cdot\va=\vp\cdot\vb=0$) and integrate out $a_0$ and $b_0$ to obtain the following dynamic theory:
\eqa
&&S=S_{int}[j_0-\rb]+
\int d^2xdt
-\frac{i}{8\pi}(\va-\vb)\times(\dot{\va}-\dot{\vb})\nonumber \\
 &&+ i\int d^2xdt\hspace{0.05in} [(\va-\vb)-\va]\cdot\vj\nonumber \\
&&Constraint:\hspace{0.05in}j_0-\rb=\frac{1}{4\pi}\vp\times(\va-\vb)\nonumber \\
&&Constraint:\hspace{0.05in}
j_0=\frac{1}{2\pi}\vp\times\va\nonumber \\
&&Constraint:\hspace{0.05in}\vp\cdot\va=\vp\cdot\vb=0.
\label{5}
\eea
Since $\vp\times\va$ is tied to electrons, Eq.(\ref{5}) suggests that $j$ is the 3-current of electrons. As the result, we identify ${\vrr_i}$ in Eq.(\ref{dii}) with the coordinates of electrons. After such identification, we can replace the $-i\va\cdot\vj$ term in the action and the constraint 
$j_0=\frac{1}{2\pi}\vp\times\va$ in Eq.(\ref{5}) by
a fermionic boundary condition on $\{\vrr_j\}$. 
After such replacement we have 
\eqa
&&S=S_{int}[j_0-\rb] + i\int d^2xdt (\va-\vb)\cdot\vj\nonumber \\
&&-\int d^2xdt
\frac{i}{8\pi}(\va-\vb)\times(\dot{\va}-\dot{\vb})
\nonumber \\
&&Constraint:\hspace{0.05in}j_0-\rb=\frac{1}{4\pi}\vp\times(\va-\vb)\nonumber \\
&&Constraint:\hspace{0.05in}\vp\cdot(\va-\vb)=0\nonumber \\
&&Constraint:\hspace{0.05in}Fermion\hspace{0.05in}boundary
\hspace{0.05in}condition\hspace{0.05in}on\hspace{0.05in}\{\vrr_j\}.
\label{6}
\eea

Because $S_{int}$ is minimized when $j_0-\rb=0$, the constraint 
in Eq.(\ref{6}) implies that it is energetically favorable for 
$\vp\times(\va-\vb)$ to vanish. One way to achieve that is for each 
quantum of $\vp\times\va$ to bind with that of $\vp\times\vb$. Due to our 
earlier charge assesment, the bound state is neutral. In addition, due to 
the fermion boundary condition on $\{\vrr_j\}$, the bound state has Fermi 
statistics. 

In configuration space, each neutral fermion is characterized by $\vrr_i$ and its dipole moment $\vec{p}_i$\cite{dipole}
The electric polarization $\vec{P}$ due to such dipoles is given by $\vec{P}\equiv\sum_j\vec{p}_j\delta(\vx-\vrr_j)$.
>From $\vec{P}$ the polarization charge and current densities are given by $J_{p0}=-\vp\cdot\vec{P},\hspace{0.05in} \vec{J}_p=\dot{\vec{P}}$. The polarization 3-current $J_p$ is related to $\gr\times(a-b)$ via
$ J_p\equiv \frac{1}{4\pi}\gr\times(a-b)$.
As the result
\eq
\va-\vb=-4\pi\zh\times\vec{P}
\label{di}
\ee

By differentiating the action in Eq.(\ref{2}) twice with respect to $A$ we obtain:
\eq
\s_{\alpha\beta}=\frac{1}{4\pi}\epsilon_{\alpha\beta}+
\s_{\alpha\beta}^{N},
\label{ppp}
\ee
where $\s_{\alpha\beta}^{N}\equiv\frac{1}{i\w} <J_{p\alpha}\cdot J_{p\beta}>
$.
Eq.(\ref{ppp}) establishes the missing link between the properties of 
electrons and neutral fermions.
Although Eq.(\ref{ppp}) is generally valid, the identification of 
$\frac{1}{4\pi}\gr\times (a-b)$ with the polarization 3-current of dipoles 
is not. This is particularly true when a strong disorder breaks the neutral 
fermions apart. 

After substituting Eqs.(\ref{dii},\ref{di}) 
into Eq.(\ref{6}) we obtain the following
first-quantized coherent-state path integral action in $\vrr$ and $\vec{\pi}$ for the neutral fermions:
\eqa
&&S=i\int dt\sum_j\vec{\pi}_j\cdot\dot{\vrr}_j+i\frac{l_0^2}{4}\sum_j\vec{\pi}_j\times\dot{\vec{\pi}_j}\nonumber \\
&&+S_{int}[\frac{l_0^2}{2}\vp\times\sum_j\vec{\pi}_j\delta(\vx-\vrr_j)],
\nonumber \\
&&Constraint:\hspace{0.05in}\sum_j\delta(\vx-\vrr_j
)-\rb=\frac{l_0^2}{2}\vp\times\sum_j\vec{\pi}_j\delta(\vx-\vrr_j)], \nonumber \\
&&Constraint:\hspace{0.05in}\vp\cdot\sum_j\vec{\pi}_j\delta(\vx-\vrr_j)=0,\nonumber \\
&&Constraint:\hspace{0.05in}Fermion\hspace{0.05in}boundary
\hspace{0.05in}condition\hspace{0.05in}on\hspace{0.05in}\{\vrr_j\}.
\label{555}
\eea
Here we have introduced $\vec{\pi}_i=-\frac{2}{l_0^2}\zh\times\vec{p}_i$. By redefining
$\vec{r}\ra\vec{r}-\frac{l_0^2}{4}\zh\times\vec{\pi}$, the above result agrees with a 
recent result obtained by Shankar and Murthy.\cite{shankar}

At this point it is useful to provide some physical arguments to rationalize Eq.(\ref{555}). The
interaction term, $S_{int}[-\vp\cdot\vec{P}]$, simply
describes the electrostatic interaction between dipoles. The term $i\int dt\sum_j\vec{\pi}_j\cdot\dot{\vrr}_j$ implying 
the conjugation between $\vrr$ and $\vec{\pi}$, is the result of the guiding-center dynamics of neutral fermions.
Finally, the term $i\frac{l_0^2}{4}\sum_j\vec{\pi}_j\times\dot{\vec{\pi}_j}$
is the Berry phase due to the rotation of the dipole.

It is interesting to note that Eq.(\ref{555}) is invariant under
$\vec{\pi}_j\ra\vec{\pi}_j+\vec{\pi}_0$ 
and $\vrr_j\ra\vrr_j+\frac{l_0^2}{2}
\zh\times\vec{\pi}_0$.\cite{pasquier} 
To deduce the dynamic consequences of
Eq.(\ref{555}), we recast 
Eq.(\ref{555}) into a form involving 
the vector fields $\vec{P}(t,\vx)$ and $\vec{j}(t,\vx)$.
\eqa
&&S=-i4\pi\int dtd^2x\vec{P}\times\vec{j}+i2\pi\int dtd^2x\vec{P}\times\dot{\vec{P}}\nonumber \\
&&+S_{int}[-\vp\cdot\vec{P}],\nonumber \\
&&Constraint:\hspace{0.05in}j_{0}-\rb=-\vp\cdot\vec{P},\nonumber \\
&&Constraint:\hspace{0.05in}\vp\times\vec{P}=0,\nonumber \\
&&Constraint:\hspace{0.05in}Fermion\hspace{0.05in}boundary
\hspace{0.05in}condition\hspace{0.05in}on\hspace{0.05in}\{\vrr_j\}.
\label{ferm}
\eea
Because $\vec{P}$ is pure logitudinal, only the transverse component of 
$\vec{j}$ couples to it. Now let us consider a long wavelength and low
frequency $P(\vq,\w)(\zh\times\hat{q})$. Through the linear responses it induces
$:j_0:=-4\pi\Pi_{lt} P$ and $j_t=-4\pi\Pi_{tt} P$. Here $\Pi_{lt}$ and 
$\Pi_{tt}$ ($l$ and $t$ stand for logitudinal and transverse respectively), yet to be
determined self-consistently, 
are electronic response functions. Combining the above results with the 
constraint $:j_0:=-\vp\cdot\vec{P}$ gives $\Pi_{lt}=\frac{iq}{4\pi}$ 
(or equivalently $\s_{xy}=1/4\pi$), while leaving $\Pi_{tt}$ undetermined. 
Therefore after integrating out the quadratic fluctuations in the 
transverse component of $\vec{j}$, the following effective theory 
governing the intra-Landau level density fluctuation is obtained:
\eq
S=S_{eff}[P]+S_{int}[-iq P(\vq,\w)],
\label{fermf}
\ee
where 
$S_{eff}[P]=\frac{1}{2}\int\frac{d\w d^2q}{(2\pi)^3}(4\pi)^2 
\Pi_{tt}(\vq,\w) P(\vq,\w)^2 + O(P^4)... .$
Eq.(\ref{fermf}) predicts an intra-Landau level density-density correlation function given by:
\eqa
&&S(\vq,\w)=\frac{q^2}{(4\pi)^2\Pi_{tt}(\vq,\w)+q^2V(\vq)}+
\rm{higher\hspace{0.05in} order\hspace{0.05in}}\nonumber \\ &&\rm{corrections}.
\label{sqw}
\eea
By making rather safe assumption on the form of $\Pi_{tt}$, we can predict 
the behavior of static compressibility. To be more specific, if we assume 
the electron has a finite diamagentic response in the DC limit (hence $\Pi_{tt}\ra |\vq|^2$), we obtain $S(\vq,0)\sim |\vq|$ or $\sim constant$ depending on whether the electron-electron interaction is Coulomb or short range respectively.
If we further assume 
that $\Pi_{tt}(\vq,\w)$
takes on a form $C_1\frac{|\w|}{q}+C_2q^2$ we find (after wick rotation) 1) $\s_{ll}(\vq,\w)=\frac{1}{i\w}\frac{\w^2}{q^2}S(\vq,\w)
\propto |\vq|$\cite{willet2,hlr}, 
2) $\int d\w S(\vq,\w)\propto |\vq|^3\ln|\vq|$,\cite{hlr} 
and 3) $S(\vq,0)\propto |\vq|$.\cite{hlr} The above results agree with 
the {\it electronic} response functions obtained in Ref.\cite{hlr}. 

In the presence of a smooth but strong\cite{str} disorder the neutral fermions will be ripped apart. In that case the constituents of neutral fermions will be attracted to the region of negative/positive potential according to its charge. If the disorder potential overwhelms electron-electron repulsion, the negatively charged anyons will form the densest possible state near the valleys of the potential. In the lowest Landau level such state is the $\nu=1$ quantum Hall state including the contribution of the $\nu=1/2$ boson Hall liquid background. Similarly, the positively charged remnants of neutral fermions accumulate 
near the hill of the potential to form the $\nu=0$ quantum Hall states (i.e. vacuum).

The above physical arguments translate into the following formal derivation. Near the valleys of the disorder potential 
$\vp\times\vb=0$, and the effective action is given by:
\eqa
S&=&S_{int}[\dva]+\int d^2xdt 
-\frac{i}{8\pi}(a-A)\cdot\gr\times (a-A)\nonumber \\
&+&\frac{i}{4\pi}a\cdot\gr\times a - i\rb (a_0-A_0).
\label{19}
\eea
We can integrate out $a$ (or equivalently letting the negatively charged constituents of neutral fermions to form the densest possible state) to obtain the following effective action
\eq
S_{eff}=S_{int}[\rb]+\int d^2xdt\{-\frac{i}{4\pi}A\cdot\gr\times A+2i\rb A_0\}.
\label{123}
\ee
Eq.(\ref{123}) describes the electromagnetic properties of a $\nu=1$ quantum Hall liquid.

Near the hills of the disorder potential, $\vp\times\va=0$ the effective theory is given by
\eqa
S&&=S_{int}[\frac{\vp\times\vb}{4p\pi}]+\int d^2xdt
\{ -\frac{i}{8\pi}(b+A)\cdot\gr\times (b+A) \nonumber \\
&&+ib\cdot j
+i\rb(b_0+A_0).\}
\label{22}
\eea
Letting $\gr\times b$ fluctuations condense 
(hence quenching the $j$ fluctuations) 
and integrating it out, we obtain the 
effective gauge action $S_{eff}=S_{int}[\rb]$. The latter is, of course, the effective action of vacuum. 

The above discussion leads to the following picture. In regions of strong disorder, the system locally realizes either the $\nu=1$ Hall state or the vacuum depending on the sign of the local potential. In the intervening space where disorder is not so strong the neutral fermion description presented above could remain valid. When the root mean square fluctuation of the disorder potential increases, the width of the intervening region becomes very narrow. When the width becomes smaller than the magnetic length, we should view the whole system as $\nu=1$ 
quantum Hall liquid percolating on a $\nu=0$ background.

The above discussions can be simply generalized to describe $\nu=1/2p$ with $p\ne 1$. Moreover, it can also be applied to describe bosons at filling factor $1/(2p+1)$.\cite{pasquier} In this paper we have ignored the possibility of neutral fermion pairing. Indeed, due to the dipole-like interaction neutral fermions can pair at very low temperatures.\cite{baskaran,pasquier}. To conclude, we 
point out the following open questions. 
1) The transport properties of neutral fermions in the presence of weak disorder is unsolved. In particular the
question concerning whether there is a metallic phase is pressing. 2) The nature of transition from the weak to 
strong disorder is unknown. For example, is there a 
quantum phase transition as a function of disorder, or just a crossover? 
The answers to these questions are extremely important
for a global understanding of the phase structure of a 2DEG 
under strong magnetic field.\cite{hlr,klz}
\\
\bf
\noindent{Acknowledgment}
\\
\rm
I thank S. Bahcall, D. Haldane, V. Pasquier, N. Read, R. Shankar, 
and especially S. Kivelson for useful discussions.


\begin{references}
\bibitem{str} The word ``strong'' means that such potential overwhelms electron-electron repulsion.
\bibitem{dipole}$\vec{p}_j$ is one half (due to the charge of $a$ and $b$
quanta) times the vector running from the $b$ quantum to the $a$ quantum.
\end{references}
\end{document}